\documentclass[lettersize,journal]{IEEEtran}
\usepackage{amsmath,amsfonts}
\usepackage{algorithm}
\usepackage[noend]{algpseudocode}
\usepackage{array}
\usepackage{textcomp}
\usepackage{stfloats}
\usepackage{url}
\usepackage{verbatim}
\usepackage{graphicx}
\usepackage{cite}
\usepackage{microtype}
\usepackage{booktabs}
\usepackage{multirow}
\usepackage{xurl}
\usepackage[hidelinks]{hyperref}
\usepackage[dvipsnames]{xcolor}
\usepackage[inline]{enumitem}
\usepackage{subcaption}
\usepackage{todonotes}

\makeatletter
\def\ps@IEEEtitlepagestyle{%
	\def\@oddfoot{\mycopyrightnotice}%
	\def\@oddhead{\hbox{}\@IEEEheaderstyle\leftmark\hfil\thepage}\relax
	\def\@evenhead{\@IEEEheaderstyle\thepage\hfil\leftmark\hbox{}}\relax
	\def\@evenfoot{}%
}
\def\mycopyrightnotice{%
	\begin{minipage}{\textwidth}
		\scriptsize
		\copyright~2025 IEEE. DOI: 10.1109/TC.2025.3551936.
		Personal use of this material is permitted. 
		Permission from IEEE must be obtained for all other uses,
		in any current or future media, including reprinting/republishing this material
		for advertising or promotional purposes, creating new collective works,
		for resale or redistribution to servers or lists, or reuse of any copyrighted
		component of this work in other works by sending a request to pubs-permissions@ieee.org.
	\end{minipage}
}
\makeatother

\begin{document}

\title{An FPGA-Based Open-Source Hardware-Software Framework for Side-Channel Security Research}

\author{Davide Zoni, Andrea Galimberti, Davide Galli
\thanks{Davide Zoni, Andrea Galimberti, and Davide Galli are with
	Dipartimento di Elettronica, Informazione e Bioingegneria,
	Politecnico di Milano, Milan 20133 Italy.
	E-mail: \{davide.zoni, andrea.galimberti, davide.galli\}@polimi.it.}
\thanks{This work was supported by the European Union’s Chips Joint Undertaking
	(Chips JU) program under grant agreement No. 101112274 (ISOLDE).
	Accepted for publication on IEEE Transactions on Computers.}
}

\markboth{}%
{Zoni \MakeLowercase{\textit{et al.}}:
An FPGA-Based Open-Source Hardware-Software Framework for Side-Channel Security Research}

\maketitle

\begin{abstract}
Attacks based on side-channel analysis~(SCA) pose a severe security threat to
modern computing platforms, further exacerbated on IoT devices by
their pervasiveness and handling of private and critical data.
Designing SCA-resistant computing platforms requires a significant additional effort in
the early stages of the IoT devices' life cycle, which is severely constrained by
strict time-to-market deadlines and tight budgets.
This manuscript introduces a hardware-software framework
meant for SCA research on FPGA targets.
It delivers an IoT-class system-on-chip~(SoC) that includes a RISC-V CPU,
provides observability and controllability through an ad-hoc debug infrastructure to
facilitate SCA attacks and evaluate the platform's security, and
streamlines the deployment of SCA countermeasures through dedicated
hardware and software features such as a DFS actuator and FreeRTOS support.
The open-source release of the framework includes the SoC, the scripts to
configure the computing platform, compile a target application, and assess the SCA security,
as well as a suite of state-of-the-art attacks and countermeasures.
The goal is to foster its adoption and novel developments in the field,
empowering designers and researchers to focus on studying SCA countermeasures and attacks
while relying on a sound and stable hardware-software platform as the foundation for their research.
\end{abstract}


\section{Introduction}
\label{sec:introduction}
\IEEEPARstart{H}{undreds} of billions of Internet of Things~(IoT) devices
are getting more and more able to autonomously make decisions thanks to
artificial intelligence.
As they get more capable, pervasive, and interconnected through 5G and 6G networks,
IoT devices continuously collect, process, and exchange sensitive and critical data,
making it paramount to consider security during their design and whole
life cycle~\cite{Hassija_2019Access}.

Relying on the theoretical security provided by traditional cryptography
solutions such as AES, RSA, and SHA-3, and by secure communication protocols
such as TLS and SSH is, however, insufficient to guarantee IoT device security.
Side-channel analysis~(SCA) attacks target information
collected from executing a specific implementation of a cryptographic
scheme or security protocol rather than flaws in their specification or
the theoretical problems at their foundation~\cite{Lou_CSUR2022}.
Such so-called side-channel information includes factors like the time taken for
cryptographic computations, power consumption, electromagnetic emissions,
and noise generated during execution.
While some fault-injection attacks disrupt the target device's normal operation,
many passive side-channel attacks, such as differential power analysis~(DPA),
remain undetectable by the system under attack.

While SCA attacks increasingly become serious security threats to IoT devices, and
in particular to those that operate in public spaces and are accessible by anyone,
the computing platforms that power these devices and that handle
sensitive and critical data are, however, often not designed to
protect against them.
Commercially available microcontroller-based system-on-chips (SoCs),
which provide computing and connectivity capabilities to IoT devices,
often execute cryptographic applications despite not being designed
with security as a primary focus, primarily due to cost-saving considerations~\cite{Barenghi_2019Access}.
Meanwhile, strict time-to-market pressures, combined with stringent constraints on
energy, power consumption, and area, along with the rapidly evolving application domain,
drive a shift towards flexible and general-purpose IoT platforms that
may support operating systems to facilitate programmability.
However, this trend towards general-purpose IoT devices results in
more complex computing platforms, making side-channel vulnerability analysis increasingly
challenging and necessitating the adoption of novel and efficient analysis frameworks.

In this context, the open literature lacks a comprehensive
hardware-software solution to address the challenges of
SCA security, particularly for modern RISC-V-based platforms,
which are promising candidates for secure-aware designs due to
their open and royalty-free instruction set architecture~(ISA).
The few RISC-V security-oriented solutions, e.g., OpenTitan~\cite{Meza_2023SP},
are not meant as general-purpose computing platforms and do not support
the development of novel SCA attack and defense methodologies,
while others focus their research effort solely on
the hardware side while disregarding the software framework~\cite{DeCnudde_2018TCHES}
or vice versa.

As new SCA attacks continue to emerge,
designing novel IoT devices that are secure against them
requires providing the system designers and security researchers with
a sound and stable hardware-software platform that lets them focus on
analyzing the security against SCA attacks and
developing countermeasures to thwart the latter.
Moreover, there is a need to not only detect the presence of vulnerabilities,
i.e., of so-called information leakage, but also to pinpoint
the specific signals in the hardware design that cause such leakage,
enabling the designers with the possibility to more easily correct
their systems and make them resistant to SCA attacks.

\subsection*{Contributions}
This manuscript introduces JARVIS,
an open-source research framework for SCA on FPGA-based IoT-class computing platforms,
that encompasses a complete SoC and a software toolchain for
SCA attacks and countermeasures, delivering a comprehensive hardware-software solution
that is, to the best of our knowledge, currently lacking in the open literature.

The framework provides an IoT-class SoC that includes
a CPU based on the RISC-V ISA,
dedicated hardware to enable the implementation of state-of-the-art SCA countermeasures,
an ad-hoc debug infrastructure to maximize the observability and controllability of
the computing platform and thus simplify the execution of SCA attacks, and
support for the open source FreeRTOS real-time operating system~(RTOS).
A complete flow encompasses the configuration of the SoC,
the execution of target applications and corresponding collection of side-channel information,
and the analysis to identify SCA vulnerabilities and leakage sources.

The goal of empowering designers and researchers to focus solely on
studying SCA countermeasures is achieved through three main contributions.
\begin{enumerate}
	\item \emph{Capability to identify leakage sources}.
	Hardware security requires not just identifying the presence of a vulnerability,
	but also pinpointing the source of such leakage.
	The computing platform has a minimal architecture to expose
	the least leakage sources and make them eventually simpler to identify,
	in addition to being easier to emulate and cheaper in terms of area and power cost,
	while a one-to-one match between the prototyped platform and
	its emulated counterpart is enforced through dedicated hardware mechanisms to
	enable the collection of the most accurate side-channel signal.
	
	\item \emph{Usability for research purposes}.
	The framework is released open source%
	\footnote{Sources available at \href{https://github.com/hardware-fab/JARVIS}
	{https://github.com/hardware-fab/JARVIS}.},
	including hardware and software,
	from the SoC to the software scripts driving the SoC configuration, compilation,
	prototype execution and emulation, oscilloscope measurement, and
	state-of-the-art SCA attacks, to foster its adoption in research settings.
	Carrying out an analysis through the JARVIS framework not only provides
	the traditional power, performance, and area (PPA) quality metrics, but
	it extends them by adding the security dimension.

	\item \emph{Complete SCA ecosystem}.
	The framework includes dedicated hardware and software support for
	state-of-the-art SCA attacks and countermeasures that can be
	employed out of the box to directly compare with solutions
	from the literature as well as evaluate and enforce the SCA security
	of the computing platform.
	The experimental evaluation in this manuscript demonstrates
	the framework's capabilities and showcases the SCA attacks and
	countermeasures included as part of its open-source release.
\end{enumerate}

\subsection*{Structure of the manuscript}
The rest of this manuscript is organized into five parts.
Section~\ref{sec:background} provides
a background on SCA attacks and countermeasures and
an overview of the state-of-the-art RISC-V computing platforms
and frameworks for SCA attacks and countermeasures.
Section~\ref{sec:infrastructure} describes the JARVIS framework
and the hardware-software infrastructure that implements it.
Section~\ref{sec:platform} details the key hardware and software aspects that
enable the SCA attack and countermeasure capabilities of the framework,
Section~\ref{sec:experimentalresults} showcases the framework capabilities
through a comprehensive experimental evaluation.
Finally, Section~\ref{sec:conclusions} draws conclusions and discusses
the future works and developments on the proposed framework.

\section{Background and Related Works}
\label{sec:background}
\subsection{Side-channel analysis attacks and countermeasures}

A vast amount of research has tackled the complementary topics of SCA attacks
and countermeasures ever since the emergence of SCA as a security threat~\cite{Kocher_1999CRYPTO}.

SCA attacks can be classified as either non-profiled or profiled.
Non-profiled ones, such as those based on
DPA~\cite{Kocher_1999CRYPTO} and
correlation power analysis~(CPA)~\cite{Brier_2004CHES},
only target side-channel information that is obtained from
the specific device under attack while it is performing a computation,
and then attempt to recover the corresponding secret key by
leveraging the statistical correlation between the measured
side-channel signal and the data being processed, exploiting
a partial knowledge of the latter.
Profiled attacks~\cite{Standaert_2009ACNS} are carried out instead by
initially making use of a replica of the device to be attacked to
identify and fine-tune the side-channel leakage model and
later employing such model to attack the actual target device,
under the assumption that the latter and its replica share
identical or at least similar leakage models~\cite{Chari_2002CHES}.
Whereas SCA attacks have traditionally employed statistical techniques,
machine- and deep-learning approaches recently emerged as
promising research avenues for more capable attacks, ranging from
identification of cryptographic operations from side-channel power traces~\cite{Chiari_2024DATE, Galli_2024ICCD}
to learning-based non-profiled~\cite{Timon_2019TCHES} and profiled~\cite{Kim_2019IACR, Galli_2024TC} attacks.

The advancements in SCA attacks are mitigated by an even larger research
effort being devoted to identifying new defense mechanisms against them.
SCA countermeasures can be mainly split into masking and hiding ones. 
Masking countermeasures split the sensitive intermediate values into
different shares that are computed independently from each other, with the goal of
minimizing the dependency of each share from the secret key~\cite{Gross_2016TIS}.
Hiding countermeasures aim instead to randomize or add noise to
the side-channel emission of the computing platform to be protected
in order to reduce the information leakage that can be exploited by an attacker.
Such countermeasures make use of techniques such as code morphing, i.e., the insertion of
random instructions in the original execution flow~\cite{Agosta_2012DAC},
clock frequency randomization~\cite{Galli_2024ICECS}, and
the concurrent computation of multiple cryptographic operations
with invalid keys~\cite{Barenghi_2020TCAD}.

\begin{figure*}[t]
	\centering
	\includegraphics[width=\textwidth]{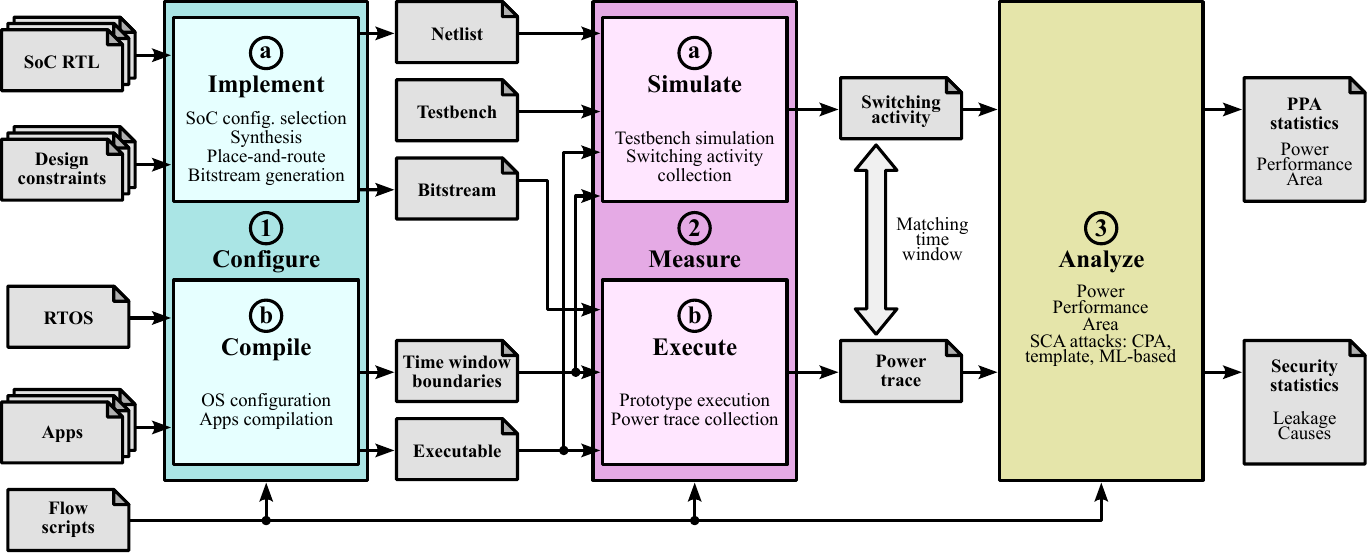}
	\caption{High-level flow of the JARVIS framework.}
	\label{fig:flow}
\end{figure*}

\subsection{State-of-the-art RISC-V computing platforms}
\label{ssec:background_sota}
RISC-V has emerged as the de-facto standard ISA for IoT-class computing platforms thanks to
its modularity, enabled by a minimal baseline instruction set whereas
separate ISA extensions are each devoted to a specific set of additional features, and
extensibility, given by the possibility of exploiting unused opcode space
to encode custom instructions and extensions according to 
the specific needs of computer designers~\cite{Waterman_TechRepUCB2016}.

On the one hand, a variety of RISC-V-based microcontroller-class SoCs is available, such as
PULPino~\cite{Traber_RISCV2016} and GAP-8~\cite{Flamand_2018ASAP},
while ultra-low-power platforms also leverage RISC-V in multi-core heterogeneous architectures,
e.g., Mr. Wolf~\cite{Pullini_2019JSSC} and HERO~\cite{Kurth_2022TPDS},
and SoCs coupling hard-core CPU clusters with programmable FPGA logic,
e.g., Microchip's PolarFire SoC~\cite{polarfire}.
Such solutions focus on delivering good performance with low energy and power consumption
but do not address security, especially SCA-related aspects.
Moreover, due to their nature, designers can not modify ASIC cores
to implement SCA countermeasures not originally included, drastically
limiting their usability for comprehensive SCA research.

On the other hand, multiple frameworks such as ESP~\cite{Mantovani_2020ICCAD} and Chipyard~\cite{Amid_2020Micro}
have emerged in the literature for the agile development of heterogeneous multi- and many-core SoCs
that feature both general-purpose CPU cores and hardware accelerators~\cite{Montanaro_2024ICCD}.
However, they focus on the modularity and composability of the obtained computing platforms and on
the fast prototyping of the latter on FPGA targets, while they do not consider instead any security aspects.

All the previously listed platforms, including microcontroller-class, ultra-low-power,
and heterogeneous multi- and many-core SoCs, are capable and efficient on the computing side.
However, their complexity hinders the possibility of simulating them in an RTL simulator,
which is crucial to detecting SCA vulnerabilities and, more importantly,
accurately identifying which signals are responsible for the information leakage,
which becomes ever more challenging to achieve as such hardware platforms keep growing in size.

Finally, the OpenTitan project~\cite{Meza_2023SP} delivers
a hardware root of trust~(RoT) centered around a RISC-V core, 
not providing instead either a complete SoC platform or
a comprehensive hardware-software framework for SCA-related research.

There also exists a variety of both RISC-V resource-constrained and application-class cores
in the open literature, e.g.,
CORE-V CV32E40P~\cite{Gautschi_TVLSI2017} and CVA6~\cite{Zaruba_TVLSI2019},
UCB's Rocket~\cite{Lee_2016Micro}, and Frontgrade Gaisler NOEL-V~\cite{Wessman_2021SCC},
that are however of limited use, when considered on their own,
in providing a full platform for SCA research.
State-of-the-art architectural simulators such as gem5~\cite{Binkert_2011SIGARCH}
and RISC-V-specific GVSoC~\cite{Bruschi_2021ICCD} operate instead at a high level of abstraction,
drastically shrinking the time required to simulate a computing platform but
conversely making them unsuitable to produce information
that can be effectively exploited by SCA attacks.

\subsection{State-of-the-art SCA frameworks}
\label{ssec:background_sca}
The literature offers few solutions that are meant for the SCA of RISC-V-based computing platforms.

lowRISC provides an open-source SCA setup for OpenTitan \cite{OpenTitanSca}
allows instantiating the OpenTitan RoT on a target NewAE board and
assessing the resistance of its cryptographic accelerators to power SCA attacks.
The OpenTitan platform occupies however a large number of FPGA resources
and requires therefore targeting large and expensive FPGAs, e.g., AMD Kintex-7 ones,
leading to non-negligible costs, and does not offer any actuators that are suitable for
protection against SCA attacks.
Moreover, the size and complexity of OpenTitan makes it complex to modify
the given design with the goal of implementing novel SCA countermeasures,
and the SCA setup \cite{OpenTitanSca} lacks the ability to easily
identify the leakage source in the microarchitectural design.

ASIC-based solutions, that include mounting on NewAE's CW308 board an FE310-G002 daughter board \cite{cw308} which features
a hard-core SiFive 32-bit RISC-V CPU but lacks any SCA protection, have two main drawbacks.
On the one hand, pinpointing the side-channel leakage within the microarchitecture is not
easily achievable due to the limited observability and closed-source nature of the computing platform.
On the other hand, the researcher can not easily modify the hardware to implement SCA countermeasures.

Finally, NewAE's CW305, CW310, and CW340 boards also allow the users to flash
their own custom bitstreams, including RISC-V-based computing platforms.
This requires however either porting a third-party computing platform to the new target or
designing a new one, and both paths notably require a large amount of effort.

Conversely, a large part of the literature in the SCA research field targets instead more outdated
hardware such as the
STM32 32-bit Arm Cortex-M~\cite{Agosta_2015TCAD} and
ATmega 8-bit AVR~\cite{Benadjila_2020JCE}
microcontrollers, highlighting the need to foster novel
SCA-related research on modern IoT-class computing platforms.

\section{Framework}
\label{sec:infrastructure}
The JARVIS framework enables thoroughly evaluating
the vulnerability to SCA attacks of cryptographic applications executed on
an IoT-class SoC that implements countermeasures through dedicated hardware support.
The possibility to accurately match the emulated and prototype execution of
the computing platform allows not only detecting information leakage,
but even more importantly identifying its sources, which is
paramount to fixing the vulnerabilities of the SoC during the design phase.
Section~\ref{ssec:infrastructure_flow} outlines the high-level flow for
the proposed framework, while Section~\ref{ssec:infrastructure_hwsw}
discusses how such flow is implemented in a hardware-software infrastructure.

\subsection{High-level flow}
\label{ssec:infrastructure_flow}
The JARVIS framework, depicted in Figure~\ref{fig:flow}, can be
described as a sequence of three main phases devoted to
\begin{enumerate*}[label=\textit{\arabic*)}]
	\item \textit{configuring} the hardware-software setup,
	\item \textit{measuring} SCA-related information from simulation and prototype execution, and
	\item \textit{analyzing} such information to detect leakage and identify its sources.
\end{enumerate*}
Remarkably, the open-source nature of the framework, as well as
the adoption of standard languages for all the inputs and its internals,
enable the users to tinker, adapt, and tailor it to their needs.
Software scripts automate the execution of each of the three phases,
which are discussed in more detail in the rest of this section.

\subsubsection{Configure}
\label{sssec:flow_configure}
The high-level flow starts with the configuration of the hardware and software parts of
the target to be simulated and prototyped.
The SoC is thus configured by the user and is implemented according to
the standard hardware design flow, while the applications and, optionally,
the RTOS are compiled to produce an executable to be run by the CPU.

\paragraph{Implement}
The framework provides the RTL description of the SoC and the design constraints files that
define the frequency of the clock signal and map the I/O to the target FPGA chip,
thus enabling the deployment on the prototype board.
The initial step requires selecting which additional components to instantiate in the SoC,
the parameterization for configurable aspects of the SoC, and which FPGA to target,
then an EDA toolchain is leveraged for
the synthesis and place-and-route of the netlist and the
generation of the corresponding FPGA bitstream.

\paragraph{Compile}
During this phase, the source code for the applications
provided by the user of the JARVIS framework is compiled,
through a compiler toolchain for RISC-V, to be executed on
the target platform, both in simulation and on the prototype.
The compilation produces an executable file for
the compiled applications and, optionally, the RTOS,
meant to be executed by the computing platform
both in simulation and on the prototype FPGA.
The boundaries for the time window of interest in
the execution of the application, e.g., the computation of
a specific cryptographic kernel of which to evaluate SCA resistance,
are also extracted from the executable file.

\subsubsection{Measure}
\label{sssec:flow_simexe}
Once the netlist and bitstream for SoC have been generated
and the applications have been compiled, the simulation
and prototype execution can be carried out to
collect the measurements meant to be used for SCA purposes.
The simulation produces switching activity statistics,
while a power trace is measured from an oscilloscope
connected to the board during the prototype execution.
The simulation and prototype execution are notably time-aligned,
i.e., synchronized, through dedicated hardware features
implemented in the target SoC platform.
Such synchronization enforces indeed a temporal match between
the switching activity obtained from simulation and
the power trace measured during the prototype execution.

\paragraph{Simulate}
The simulation, in a SystemVerilog testbench provided as part of the framework, of
the post-place-and-route netlist of the SoC running the compiled application executable produces
the switching activity of the internal signals of the SoC in the time window of interest.

\paragraph{Execute}
After flashing the bitstream to the FPGA mounted on the prototype board,
the SoC is fed the application binary to be loaded into memory
and then executes it.
An oscilloscope collects the power trace measurement for the execution
of the target application on the prototype board within the time window of
interest, matching the one employed in the corresponding simulation.

\subsubsection{Analyze}
\label{sssec:flow_analyze}
The final phase of the JARVIS framework
foresees the analysis of the measurements from the previous
phase, i.e., the switching activity of the simulation and
the power consumption trace of the board execution, both
corresponding to the time window of interest
and synchronized with each other.
Two degrees of analysis are supported by the synchronization and
alignment between the switching activity and the power trace.
On the one hand, SCA leakage and the corresponding
time instant are detected from the physical power traces.
On the other hand, and more importantly, such alignment helps to
identify the specific signals of the design under analysis which
are responsible for the SCA leakage, providing an in-depth understanding of
the leakage behavior that drastically eases the subsequent ad-hoc design of effective countermeasures.
A set of state-of-the-art SCA techniques,
ranging from CPA and template to ML-based ones, is
therefore provided to evaluate the SCA security of
the platform and the application executed on top of it.
Remarkably, the analysis phase also produces the traditional
power, performance, and area~(PPA) metrics.

\begin{figure*}[t]
	\centering
	\includegraphics[width=\textwidth]{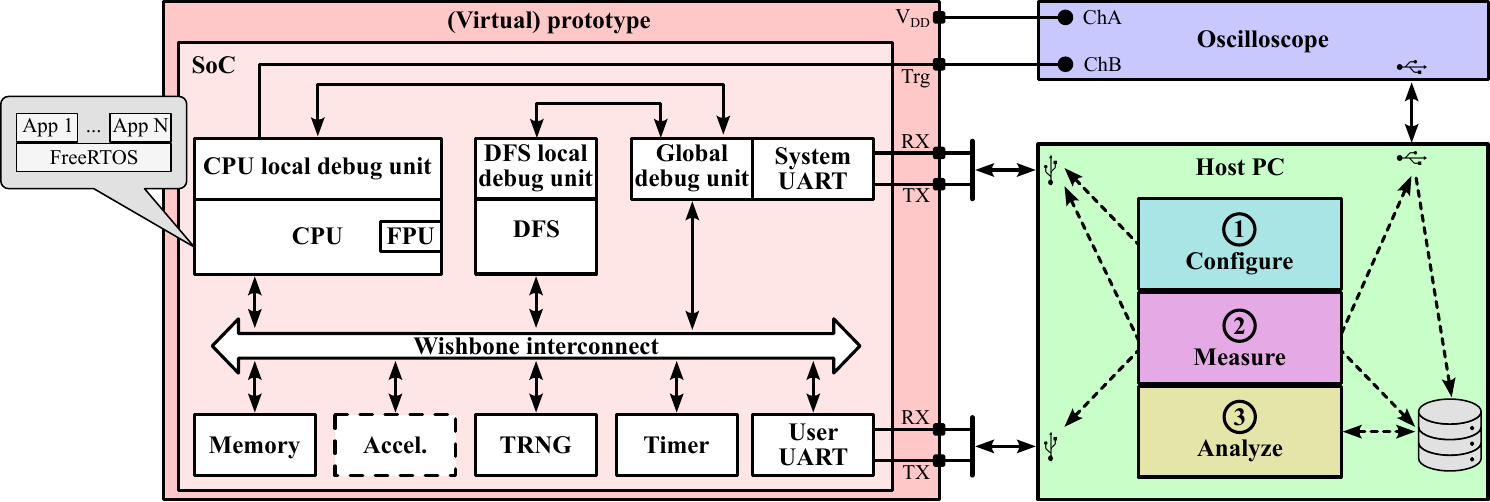}
	\caption{Hardware-software infrastructure that implements the JARVIS framework.}
	\label{fig:soc}
\end{figure*}

\subsection{Hardware-software infrastructure}
\label{ssec:infrastructure_hwsw}
The hardware-software infrastructure depicted in Figure~\ref{fig:soc}
realizes the high-level flow described in Section~\ref{ssec:infrastructure_flow}
by means of three main components, namely
\begin{enumerate*}[label=\textit{\arabic*)}]
	\item a (virtual) prototype,
	\item an oscilloscope, and 
	\item a host PC.
\end{enumerate*}
This part discusses in detail the three components and how
they interact with each other to deliver the JARVIS framework.

\subsubsection{(Virtual) prototype}
\label{sssec:infr_hwsw_proto}
The computing platform at the foundation of the proposed
SCA research framework implements an SoC architecture that can
notably be instantiated as a prototype on FPGA,
to collect the power trace measurements from an oscilloscope,
as well as emulated as a \textit{virtual} prototype in a RTL simulator
that instantiates its post-place-and-route netlist in a testbench,
to generate the matching VCD switching activity.
The correspondence between the prototype and its virtual counterpart,
in particular in how they execute a target application, is enforced
by dedicated hardware mechanisms and it is the crucial aspect that
enables employing our framework for SCA analysis.

The SoC architecture is built around a Wishbone interconnect.
It comprises a CPU, a dynamic frequency scaling~(DFS) actuator, and a global debug unit as its masters
and a memory, a user UART, a true random number generator~(TRNG), and a timer as its bus slaves.

The single-core, in-order, five-stage pipelined, 32-bit RISC-V CPU implements
the base integer~(I) and integer multiplication and division~(M) RISC-V 32-bit ISA extensions~\cite{Scotti_2019JLPEA}.
Supporting only I and M extensions, i.e., the bare minimum extensions
for an IoT-class CPU, provides the simplest and most observable setup,
making it easier to carry out the side-channel analysis.
A floating-point unit~(FPU)~\cite{Zoni_2021SUSCOM,Zoni_2022JSA,Denisov_2024JSA} can be
optionally instantiated as a functional unit of the CPU to expand
the computational capabilities of the SoC, enabling the execution of
floating-point-intensive applications without affecting
the overall side-channel resistance of the computing platform.
The interconnect consists of two separate 64-bit Wishbone data and instruction buses,
both supporting single read and write transactions as well as burst ones,
according to a modified Harvard architecture to minimize contention.
The DFS actuator enables changing at run time the frequency of
the clock signal fed to the CPU, whereas
the rest of the SoC operates at a fixed clock frequency.
A global debug unit and local debug units connected to the former
in a point-to-point fashion compose the debug infrastructure of the SoC.
The global debug unit interacts with the host PC through the system UART
while two different channels support its communication with the rest of the SoC.
Memory location of the memory-mapped bus slaves are accessed through Wishbone reads and writes,
while dedicated lines connect the global debug unit to the two local ones that
manage the CPU and DFS actuator, respectively.

The main memory, making use of the block RAM~(BRAM) resources available on AMD FPGAs,
is instantiated as a slave on the Wishbone data bus.
The bus slaves include a TRNG, that produces a sequence of random bits
meant to be used in cryptography and SCA countermeasure tasks,
a timer, which enables support for FreeRTOS,
and a user UART, that provides an I/O interface for the application.
Hardware accelerators exposing a Wishbone interface can also notably be added as
memory-mapped bus slaves to the SoC to extend its capabilities.

The implementation phase of the proposed flow includes
the selection of which parts of the SoC to instantiate,
e.g., the optional DFS actuator, TRNG, and timer, and
the configuration of its parametric features,
e.g., the baud rate of the UART interface, the width of the instruction and data buses,
and the branch prediction scheme of the CPU.

In addition to the UART I/O interfaces, the prototype exposes
the FPGA voltage~(\texttt{V\textsubscript{DD}} in Figure~\ref{fig:soc}),
which is related to the power consumption of the whole FPGA chip,
and a trigger signal~(\texttt{Trg}), that is driven by the local debug unit of the CPU.

\subsubsection{Oscilloscope}
\label{sssec:infr_hwsw_oscilloscope}
Power measurements from the prototype execution are carried out through
an oscilloscope with two analog channels and a frequency bandwidth that is
sufficient to collect samples from the target platform under measurement
without aliasing.
The oscilloscope is connected to the prototype board, with
an analog channel measuring the voltage of the FPGA and
the other one monitoring a signal meant to trigger the data acquisition on the former.
It is managed through a USB interface by the host PC, which takes care of its configuration and
receives the data samples measured from the board on both analog channels.

\begin{figure*}[t]
	\centering
	\includegraphics[width=\textwidth]{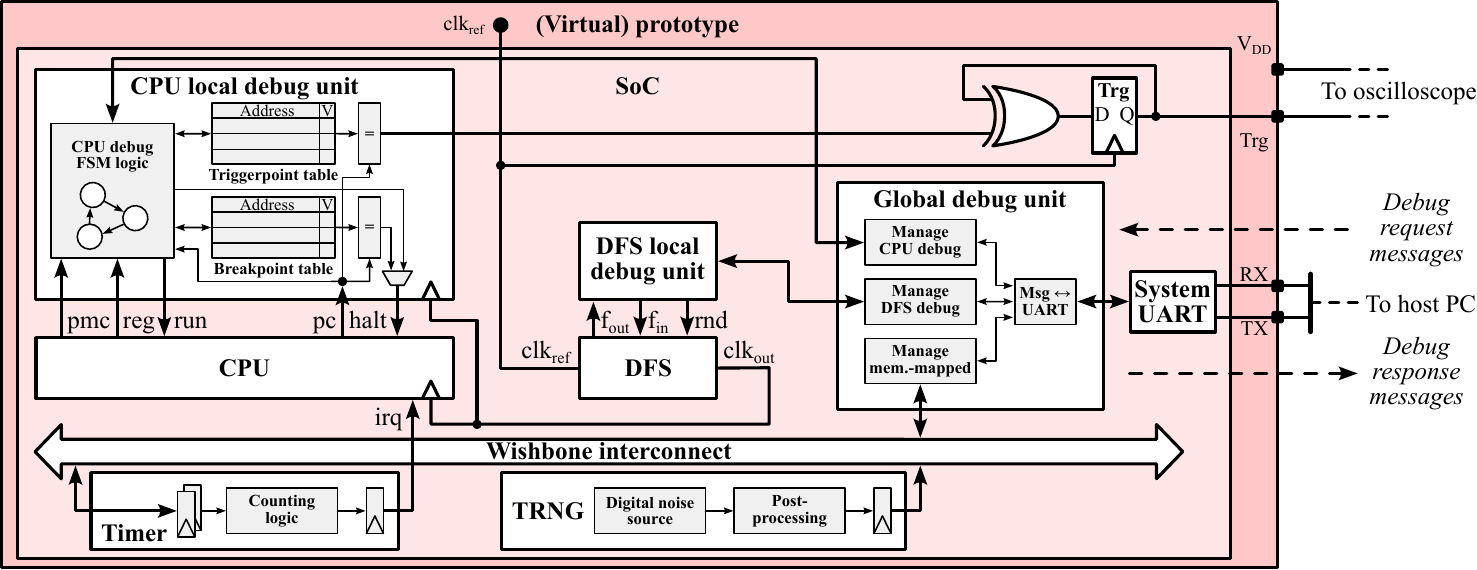}
	\caption{Detailed microarchitecture of the debug subsystem, timer, and TRNG.}
	\label{fig:debug}
\end{figure*}

\subsubsection{Host PC}
\label{sssec:infr_hwsw_host}
The host PC drives the whole JARVIS framework through the flow software scripts,
which manage the interaction with both hardware devices, namely
the prototype board and the oscilloscope, and software tools, such as
the EDA and compilation toolchains and state-of-the-art SCA attack scripts.
Matching the high-level flow previously described in Section~\ref{ssec:infrastructure_flow},
we can identify three main top-level scripts that drive
the corresponding phases of the framework.

\paragraph{Configure}
The configuration script leverages an EDA toolchain for
the synthesis and place-and-route and bitstream generation of the SoC,
previously configured according to the needs of the user,
also taking as its inputs the XDC constraint files
that include information about the internal clock signals
as well the I/O mapping on the target FPGA chip.
A RISC-V compiler toolchain is employed instead to compile
the applications and, optionally, the RTOS.
The compilation of the sources can thus include the sources for FreeRTOS, 
when targeting the execution of applications on top of the RTOS,
or not, when the goal is instead bare-metal execution.
A linker script and startup routines to be executed before
the program's main function are included to support the latter case.
The compilation process produces an executable file, that can be loaded
into the memory of the computing platform to be executed
both in its simulation and on the prototype FPGA.
Boundaries for a time window of interest are obtained from the executable to
enable accurately matching, also from a temporal standpoint,
the simulation of the computing platform with its prototype execution.

\paragraph{Measure}
The simulation leverages an RTL simulator to simulate
the post-place-and-route netlist of the SoC inside
a SystemVerilog testbench provided as part of the framework.
The testbench loads the VMEM for the target application
into the memory of the target SoC and drives the execution of
the application according to the time window boundaries
extracted from its ELF file at the previous phase.
The simulation outputs a value change dump~(VCD) that contains
all the switching activity of the internal signals of the SoC
corresponding to the execution of the application within
the time window of interest
The prototype execution requires, after flashing the bitstream to the FPGA mounted on the prototype board,
feeding the application binary to the SoC to be loaded into its memory.
The flow script properly drives the board execution through the SoC debug interface,
matching the behavior of the SystemVerilog testbench, and manages
the oscilloscope connected to the board to collect the power trace
of the execution.
The VCD from the simulation and the power trace from the prototype execution are
saved to a data storage device, so that they can be later retrieved for
performing the various analyses.

\paragraph{Analyze}
The VCD obtained from the simulation and the power traces collected from
the prototype execution, both corresponding to the exact same time window,
are retrieved from the storage where they were saved to analyze them through
a set of state-of-the-art SCA techniques, including CPA, template, and ML-based attacks,
to detect whether there was any cryptographic information leakage and
also identify the eventual sources of the latter.
In addition to the SCA security statistics, the analysis phase
outputs the traditional PPA metrics leveraging
the power consumption measured from the board execution,
the cycle-granularity latency collected from simulation, and
the resource utilization reports from netlist implementation.
The outputs of the computation can also be obtained
through the debug infrastructure to further check
the correct functioning of the system.

\section{Microarchitecture}
\label{sec:platform}
Dedicated hardware and software mechanisms enable
the SCA attack and countermeasure capabilities of
the proposed framework. 
The debug subsystem, with its breakpointing and ad-hoc triggering mechanism,
ensures maximum observability and controllability of the computing platform,
thus allowing the collection of accurate side-channel information that
can then be exploited by SCA attacks of different kinds.
A TRNG, a DFS actuator, and a timer can instead be optionally instantiated
in the SoC to enable support for a variety of state-of-the-art SCA countermeasures
ranging from masking to hiding ones.
The rest of this section delivers a detailed discussion of
the microarchitecture of the debug subsystem, TRNG, DFS actuator, and timer.

\begin{figure*}
	\begin{minipage}[b]{0.49\textwidth}
		\centering
		\begin{subfigure}[t]{\textwidth}
			\centering
			\includegraphics[width=\textwidth]{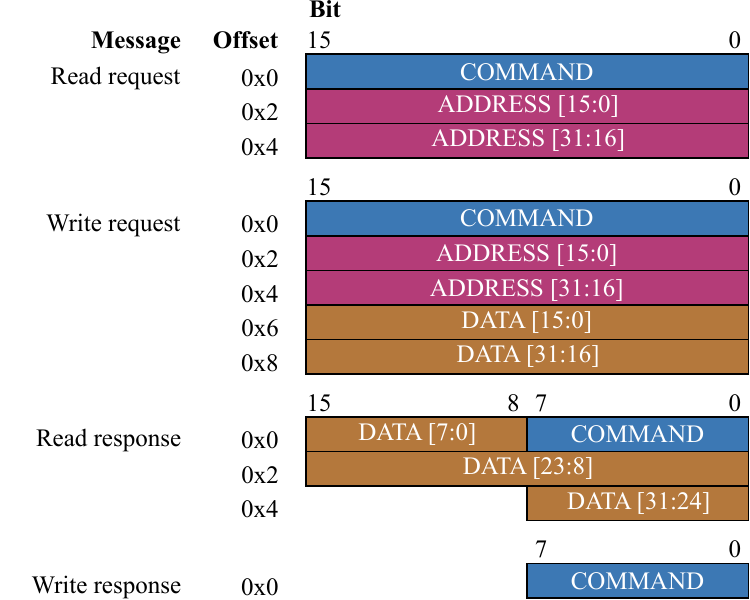}
			\caption{Debug messages}
			\label{sfig:messages}
		\vspace{0.6cm}
		\end{subfigure}
		\vfill
		\begin{subfigure}[b]{\textwidth}
			\centering
			\includegraphics[width=\textwidth]{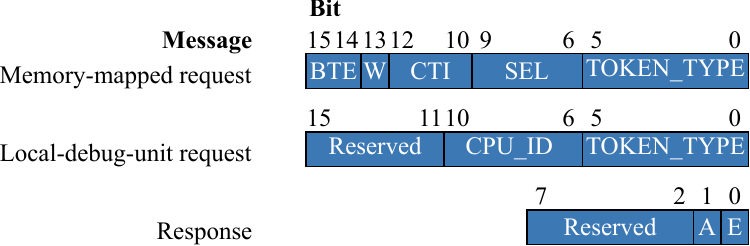}
			\caption{Command field}
			\label{sfig:commands}
		\end{subfigure}
	\end{minipage}
	\hfill
	\begin{minipage}[b]{0.49\textwidth}
		\begin{subfigure}[b]{\textwidth}
       		\centering
       		\footnotesize
\begin{tabular}{llll}
	\toprule
	\textbf{TOKEN\_TYPE}                          & \textbf{R/W}          & \textbf{Destination}        & \textbf{Group}                 \\ \midrule
	\multirow{2}{*}{INVALID}                      & \multirow{2}{*}{Read} & Memory-                     & \multirow{2}{*}{Invalid}       \\
	                                              &                       & mapped                      &                                \\ \midrule
	MMAP\_READ                                    & Read                  & Memory-                     & Memory-                        \\
	MMAP\_WRITE                                   & Write                 & mapped                      & mapped                         \\ \midrule
	GPR\_INT32\_READ                              & Read                  & \multirow{3}{*}{CPU local}  & \multirow{3}{*}{CPU}           \\
	GPR\_INT32\_WRITE                             & Write                 & \multirow{3}{*}{debug unit} & \multirow{3}{*}{registers}     \\
	GPR\_FPU32\_READ                              & Read                  &                             &                                \\
	GPR\_FPU32\_WRITE                             & Write                 &                             &                                \\ \midrule
	HALT\_CPU                                     & Read                  & \multirow{4}{*}{CPU local}  & \multirow{4}{*}{CPU reset/}    \\
	RUN\_CPU                                      & Read                  & \multirow{4}{*}{debug unit} & \multirow{4}{*}{resume/halt}   \\
	RST\_CPU                                      & Read                  &                             &                                \\
	GET\_DULOCAL\_STATE                           & Read                  &                             &                                \\
	GET\_CPU\_PC                                  & Read                  &                             &                                \\ \midrule
	ADVANCE\_ONE\_STEP                            & Read                  & \multirow{1}{*}{CPU local}  & CPU                            \\
	ECHO\_FRONTEND                                & Write                 & \multirow{1}{*}{debug unit} & stepping                       \\ \midrule
	GET\_LOW\_CYCLECNT                            & Read                  & \multirow{3}{*}{CPU local}  & \multirow{3}{*}{Performance}   \\
	GET\_HIGH\_CYCLECNT                           & Read                  & \multirow{3}{*}{debug unit} & \multirow{3}{*}{counters}      \\
	GET\_LOW\_INSTRCNT                            & Read                  &                             &                                \\
	GET\_HIGH\_INSTRCNT                           & Read                  &                             &                                \\ \midrule
	SET\_BRKPNT\_CPU                              & Write                 & \multirow{3}{*}{CPU local}  & \multirow{3}{*}{Breakpoints}   \\
	GET\_BRKPNT\_CPU                              & Read                  & \multirow{3}{*}{debug unit} & \multirow{3}{*}{configuration} \\
	RM\_BRKPNT\_CPU                               & Write                 &                             &                                \\
	GET\_NUM\_BRKPNT\_CPU                         & Read                  &                             &                                \\ \midrule
	SET\_TRGPNT\_CPU                              & Write                 & \multirow{3}{*}{CPU local}  & \multirow{3}{*}{Triggerpoints} \\
	GET\_TRGPNT\_CPU                              & Read                  & \multirow{3}{*}{debug unit} & \multirow{3}{*}{configuration} \\
	RM\_TRGPNT\_CPU                               & Write                 &                             &                                \\
	GET\_NUM\_TRGPNT\_CPU                         & Read                  &                             &                                \\ \midrule
	SET\_FREQ\_DFS                                & Write                 & \multirow{2}{*}{DFS local}  & \multirow{2}{*}{DFS}           \\
	GET\_FREQ\_DFS                                & Read                  & \multirow{2}{*}{debug unit} & \multirow{2}{*}{configuration} \\
	RND\_FREQ\_DFS                                & Write                 &                             &                                \\ \bottomrule
\end{tabular}
			\caption{Token types}
			\label{stab:types}
		\end{subfigure}
	\end{minipage}
	\caption{Debug messages supported by the proposed framework:
		(a) structure and width of the debug messages,
		(b) encoding of the command field,
		(c) token types of the request messages.
		Legend:
		\textbf{BTE} burst type extension,
		\textbf{W} write enable (0: read, 1: write),
		\textbf{CTI} cycle type identifier (for burst mode),
		\textbf{SEL} select,
		\textbf{TOKEN\_TYPE} request message type,
		\textbf{Reserved} reserved for future use,
		\textbf{CPU\_ID} identifier for target local debug unit,
		\textbf{A} ack,
		\textbf{E} error;
		\textbf{BTE}, \textbf{W}, \textbf{CTI}, and \textbf{SEL}
		refer to Wishbone.
	}
	\label{fig:messages}
\end{figure*}

\subsection{Debug subsystem}
\label{ssec:platform_debug}
The framework provides maximum observability and controllability of
the computing platform through the hardware support for breakpoints and triggerpoints
as well as a dedicated debug infrastructure that exposes a GDB-like debug interface,
thus enabling the collection of accurate side-channel information that can then be
targeted by SCA attacks of different kinds.
This part discusses first the global-local debug microarchitecture,
then focuses on the breakpointing and triggering mechanisms, and
finally provides an overview of the debug capabilities that can
be exploited by the external host PC.

\subsubsection{Global and local debug units}
The debug subsystem enables controlling and observing the whole SoC through a message-based protocol.
It is composed of a global debug unit and of a number of local debug units.
The global one, that acts as a master on the Wishbone bus, receives
debug messages through a system UART interface and accordingly communicates
both with the other bus masters, through point-to-point connections to
local debug units that act as adapters to interface with such masters, and
with the bus slaves, with whom it interacts directly through the bus
by exploiting their memory-mapped nature.

Two local debug units are instantiated for the CPU core and
for the DFS actuator, respectively,
to act as adapters between the global debug unit and the two bus masters.
The local debug unit interface with the global one is common to all the local debug units,
while the interface with the CPU or DFS actuator is custom tailored to
the specific interactions that are implemented with such module.
The CPU local debug unit can access the program counter~(PC), registers, and
performance monitoring counters~(PMCs) of the CPU, as well as halt,
reset, and restart it and advance its execution by a single step.
The DFS one can set a new target clock frequency for the DFS actuator,
get the frequency of the current clock signal generated and
configure the DFS to randomly switch the frequency of the clock signal.
Figure~\ref{fig:debug} depicts the microarchitecture of the debug subsystem
of the SoC, highlighting the global debug unit, the CPU and DFS local debug units,
and their interactions with each other as well as with the CPU and the DFS actuator.

\subsubsection{Breakpoints and triggerpoints}
The CPU local debug unit supports, through its interaction with the CPU,
a traditional breakpointing system coupled with an ad-hoc triggering one.
Breakpoints enable halting the CPU when the PC matches
their corresponding addresses, while triggerpoints, an ad-hoc variant of breakpoints,
are meant to toggle a trigger signal that drives the data acquisition from the oscilloscope
and do not instead halt the CPU, which is kept regularly running.
The local debug unit for the CPU contains two tables for the breakpoint and
triggerpoint addresses, respectively, as shown in Figure~\ref{fig:debug}.
Both tables include a configurable number of entries, each
composed of a 32-bit address that corresponds to an instruction
in the target application and a 1-bit flag that signals
its validity.

The valid addresses in the breakpoint table are constantly checked against
the current PC of the CPU, whose value is passed to the CPU local debug unit.
Whenever there is a match between a valid breakpoint address and the current PC,
the CPU is halted until it is resumed through a dedicate debug command.
The triggerpoint table entries are similarly compared against the PC,
but a match between the latter and a triggerpoint address produces
a notably different effect.
Rather than halting the CPU as in a traditional breakpoint fashion,
reaching a triggerpoint toggles a 1-bit signal that is mapped on
an I/O pin of the prototype board to be used as
a trigger signal for an oscilloscope.

\subsubsection{Debug messages}
The debug infrastructure exposes a request-response protocol
to interface with the SoC through the system UART.
The communication is indeed carried out as
a sequence of request and response messages.
Notably, no new request message can be issued until
the previous request has been completed and
the corresponding response message has been sent back.
Figure~\ref{fig:messages} summarizes the width, structure, and information
encoded in the various messages depending on whether they are request or response and
read or write and on whether they are intended for memory-mapped recipients or local debug units.

Request messages can be of four types, depending on whether
they correspond to read or write actions and whether
their recipients are memory-mapped devices
or bus masters.
All request messages are composed of a 16-bit command
and a 32-bit, while write requests also comprise a
32-bit data, as shown in Figure~\ref{sfig:messages}.
The command field encodes different information depending on whether
the debug message targets a memory-mapped peripheral or a local debug unit,
as depicted in Figure~\ref{sfig:commands}.
Debug request messages that can be sent to the SoC through the system UART include,
as listed in Figure~\ref{stab:types},
\begin{enumerate*}[label=\textit{\roman*)}]
	\item memory-mapped reads and writes,
	\item register reads and writes,
	\item commands to reset, resume, and halt the CPU,
	advance its execution by a single step, and
	get its current state and program counter,
	\item commands to retrieve performance monitoring counters
	\item commands to set, get, and remove both the breakpoints and the triggerpoints, and
	\item commands to set and get the target clock frequency for the DFS actuator and
	configure it to randomly switch clock frequencies.
\end{enumerate*}

Response messages can be instead of only two types, i.e., either read or write ones.
The former include data as part of the response, while the latter only
acknowledge the completion of the requested operation or the occurrence of an error.

\subsection{TRNG}
\label{ssec:platform_trng}
The TRNG is a Wishbone slave peripheral
that exposes a set of 32 random bits through a memory-mapped register.
The TRNG can be configured in the architecture of the digital noise sources and
post-processing methods that compose it to obtain different results in terms of
FPGA resource utilization, throughput, and security.
A digital noise source produces the actual entropy underlying
the random number generation, while a post-processing method
improves the statistical and security properties of the TRNG.
Three digital noise sources and three post-processing methods from the literature
are provided with the proposed framework to implement the TRNG component.
The digital noise sources that can be instantiated in the TRNG module
are NLFIRO, PLL-based, and edge-sampling ones, and they can be coupled with
XOR, Von Neumann, and LFSR post-processing methods~\cite{Galli_2022SAMOS}.
The random output produced by the TRNG component is periodically refreshed and
exposed by a memory-mapped register that can be read through Wishbone.

\begin{figure}[t]
	\centering
	\includegraphics[width=\columnwidth]{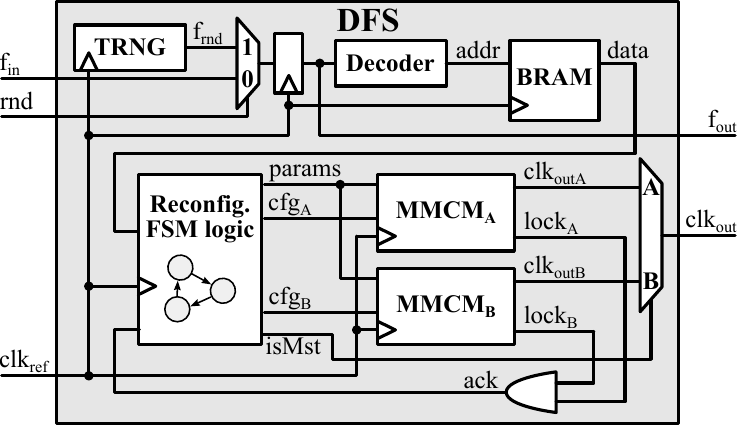}
	\caption{Detailed microarchitecture of the DFS actuator.}
	\label{fig:dfs}
\end{figure}

\subsection{DFS actuator}
\label{ssec:platform_dfs}
The DFS actuator, whose microarchitecture is depicted in Figure~\ref{fig:dfs},
leverages two mixed-mode clock manager~(MMCM) components
to provide a glitch- and latency-free switching
between different clock frequencies at run time~\cite{Galli_2024ICECS}.
It receives as its inputs a target clock frequency~(\texttt{f\textsubscript{in}} in Figure~\ref{fig:dfs}),
a 1-bit flag to enable random DFS~(\texttt{rnd}), and
a reference clock signal~(\texttt{clk\textsubscript{ref}}), and
it outputs the generated clock signal~(\texttt{clk\textsubscript{out}})
and the current clock frequency~(\texttt{f\textsubscript{out}}).
The \texttt{rnd} flag selects the actual target clock frequency for the DFS reconfiguration
between the one received through the \texttt{f\textsubscript{in}} input by the DFS local debug unit and
a random value~\texttt{f\textsubscript{rnd}} output by an internal TRNG with
similar to the one in Section~\ref{ssec:platform_trng}.
Such target clock frequency, registered to be output to the DFS local debug unit and readable
through a dedicated debug command, is decoded and used as an address to select a corresponding
set of MMCM configuration parameters from the \texttt{BRAM} memory.

Notably, the clock signal output by the MMCMs of an AMD FPGA
remains low during its reconfiguration, thus causing
a clock-gating effect on the computing platform.
The proposed DFS actuator avoids such negative effect by employing two MMCM
components~(\texttt{MMCM\textsubscript{A}} and \texttt{MMCM\textsubscript{B}} in Figure~\ref{fig:dfs}),
among which the FSM logic selects a master and a slave.
The master MMCM keeps generating the output clock signal while the slave one is
under reconfiguration, after which their roles are swapped.
An \texttt{ack} signal monitors the locking status~(\texttt{lock\textsubscript{i}}) of
the two MMCMs to prevent triggering a new reconfiguration through
the \texttt{cfg\textsubscript{i}} flag while
the previous one is still in progress.
The \texttt{clk\textsubscript{out}} clock output by the DFS actuator
is selected through the \texttt{isMst} 1-bit flag as the one output by
the current master between the \texttt{clk\textsubscript{outA}} and
\texttt{clk\textsubscript{outB}} clock signals output by the two MMCMs, respectively.

The flexible design of the DFS architecture supports
up to 1024 different clock frequency configurations.
The \texttt{BRAM} memory can indeed store up to 1024 user-defined
sets of MMCM configuration parameters, with a minimum step of $0.125$MHz and
lowest and highest achievable clock frequencies of $5$MHz and $800$MHz,
respectively, due to MMCM feasibility constraints.
Random reconfiguration is activated through a debug
request~(\texttt{RND\_FREQ\_DFS} in Figure~\ref{stab:types}),
which is dispatched to the DFS local debug unit that
in turn asserts the \texttt{rnd} input flag.
When \texttt{rnd} is set to 1, the MMCMs keep indeed reconfiguring to
new clock frequencies as soon as the previous reconfiguration has been completed,
i.e., the \texttt{ack} signal to the FSM is asserted.

\subsection{Timer and FreeRTOS support}
\label{ssec:platform_timer}
The timer is a memory-mapped peripheral that exposes
two 64-bit registers which can be accessed through
read and write requests on the Wishbone interconnect.
The two registers correspond to the current time,
which gets increased by 1 at each clock cycle, and to
the time threshold.
Once the current time is greater than the time threshold,
an interrupt request to the CPU is raised by setting to 1
the content of a 1-bit register.

Instantiating a timer in the SoC notably enables executing
the FreeRTOS real-time operating system and thus providing
support for implementing coarse-grained multithreading.

\section{Experimental Evaluation}
\label{sec:experimentalresults}
The experimental campaign aims to showcase
how the JARVIS framework delivers an effective research platform
for SCA attacks and countermeasures on FPGA targets.
Section~\ref{ssec:exp_hwsw} lists the hardware and software required by the framework,
Section~\ref{ssec:exp_setup} details the specific setup employed for the experimental evaluation,
and Section~\ref{ssec:exp_analysis} delivers an analysis of
the SCA techniques provided out of the box.

\subsection{Hardware and software requirements}
\label{ssec:exp_hwsw}
The framework makes use of widely available devices and tools,
including the development board and FPGA chips for prototyping,
the digital oscilloscope for power trace measurement,
and the software required for the various phases of the flow,
ranging from EDA synthesis and place-and-route
to compilation and SCA resistance evaluation.

\subsubsection{Prototype}
XDC design constraint files are provided for prototyping on
the NewAE Technology CW305 Artix FPGA Target board, specifically
designed for power analysis and fault injection attacks against
hardware cryptographic functions%
\footnote{More information on NewAE Technology CW305 Artix FPGA Target
available at \href{https://rtfm.newae.com/Targets/CW305\%20Artix\%20FPGA}
{https://rtfm.newae.com/Targets/CW305\%20Artix\%20FPGA}.}.
The CW305 board mounts a BGA socket that can accommodate any chip
from the AMD Artix-7 mid-range FPGA family, which is the most widely
adopted in academic and industrial research, including the security and
cryptography fields~\cite{NIST_IR8413}.
The proposed SoC fits into AMD Artix-7 50 chips, which pack
32600 look-up tables~(LUT),
65200 flip-flops~(FF),
120 digital signal processing~(DSP) blocks, and
75 36kb blocks of block RAM~(BRAM), and
larger ones from the same family.

\subsubsection{Oscilloscope}
Power measurements are carried out through an oscilloscope from
the Pico Technology PicoScope 5000 Series family,
which features a 200MHz maximum bandwidth
and can collect up to 1 billion samples per second at
a resolution ranging from 8 to 16 bits%
\footnote{More information on Pico Technology PicoScope 5000 Series available at
\href{https://www.picotech.com/oscilloscope/5000/picoscope-5000-specifications}
{https://www.picotech.com/oscilloscope/5000/picoscope-5000-specifications}.}.
The oscilloscope has its A analog channel connected to
the board through a coaxial SMA connector to measure the FPGA voltage,
while its B analog channel is connected to a general-purpose I/O pin
that outputs the trigger signal which marks the start and the end of the computation.

\subsubsection{Host PC}
The host PC must be able to execute the AMD Vivado toolchain for
the synthesis, place-and-route, and bitstream generation
targeting AMD Artix-7 FPGAs, as well as for
the simulation and switching activity VCD collection.
The host PC is also in charge of the compilation of the application
sources, optionally together with the FreeRTOS ones, through
a C compiler toolchain for RISC-V targets.
Drivers and software for the CW305 board and PicoScope 5000 oscilloscope,
connected to the host PC through USB interfaces, are required to
drive the prototype execution and power trace collection.
All the VCD, power trace, area, and execution time data is stored
permanently on a disk mounted on the host PC to be used in the analysis phase.
Support for Python 3.9 or higher is necessary to run the software scripts that
manage the whole framework as well as the scripts that
implement state-of-the-art SCA attack techniques.

\subsection{Experimental setup}
\label{ssec:exp_setup}
\subsubsection{Software setup}
The framework is run in Ubuntu 22.04.3 LTS on a host PC
that features an Intel i7-10700 CPU and a 64GB DDR4 memory.
The host PC includes ChipWhisperer 5.1.0 software and
PicoScope 7 software and drivers to manage
the CW305 board and the oscilloscope, respectively.
Applications are compiled with GCC 11.4.0 to be executed
on the SoC either bare-metal or on top of FreeRTOS 11.0.1.
AMD Vivado ML 2023.1 is employed for the RTL synthesis,
place-and-route, bitstream generation, and simulation.

\begin{table}[t]
	\centering
	\caption{Breakdown of the SoC's FPGA resource utilization.}
	\begin{tabular}{lcccc}
		\toprule
		                     &  \multicolumn{4}{c}{\textbf{FPGA resource utilization}}   \\ \cmidrule(lr){2-5}
		\textbf{Component}   & \textbf{LUT} & \textbf{FF} & \textbf{DSP} & \textbf{BRAM} \\ \midrule
		CPU                  &     4386     &     3192    &      4       &       0       \\
		DFS                  &     822      &     225     &      0       &       1.5     \\
		Global debug unit    &     286      &     248     &      0       &       0       \\
		CPU local debug unit &     875      &     587     &      0       &       0       \\
		DFS local debug unit &     19       &     29      &      0       &       0       \\
		Memory               &     206      &     138     &      0       &       64      \\
		TRNG                 &     2486     &     1252    &      0       &       0.5     \\
		Timer                &     144      &     163     &      0       &       0       \\
		System UART          &     365      &     292     &      0       &       0       \\
		User UART            &     359      &     284     &      0       &       0       \\ \cmidrule(lr){2-5}
		\textbf{Overall SoC} &     10667    &     7563    &      4       &       66      \\ \bottomrule
	\end{tabular}
	\label{tab:area}
\end{table}

\subsubsection{Hardware setup}
The FPGA target in the experimental evaluation is
an Artix-7 100~(xc7a100tftg256-1) chip mounted on a CW305 board.
The oscilloscope is a PicoScope 5244D, that features
two analog channels and a 200MHz bandwidth.
The SoC to be prototyped on FPGA and simulated is configured with
a CPU that implements the RV32IM instruction set,
a 256kB main memory, and
a TRNG, a DFS actuator, and a timer that enable support for
all the included state-of-the-art SCA countermeasures.
Table~\ref{tab:area} lists the FPGA resource utilization resulting from
synthesis and place-and-route targeting 100 MHz and 50MHz clock frequencies
for the CPU, driven by the DFS actuator, and the rest of the SoC.

\subsubsection{Breakpoints and triggerpoints}
Two triggerpoint addresses are assigned, respectively, the addresses
of instructions in the target application between which to
acquire the power measurement from the oscilloscope.
In particular, reaching the start triggerpoint address raises
a trigger signal mapped on an I/O pin of the prototype board, and
such trigger signal stays high until the PC holds a value matching
the end triggerpoint, thus toggling the trigger signal down to 0.
Two breakpoints are also leveraged, coupled with the triggerpoints,
to manage the oscilloscope's behavior.
A breakpoint is configured to an address that is reached before
the start triggerpoint one is employed to interrupt the CPU computation
in order to get the oscilloscope ready to perform the acquisition,
while another breakpoint is instead assigned to an address being encountered
after the acquisition time window to interrupt again the CPU computation and
dump the collected data that is stored in the oscilloscope's buffer. 

Such combined usage of the breakpoints and triggerpoints in the prototype execution
scenario is accurately replicated in the simulation one for two main purposes.
On the one hand, it enables a perfect match from the temporal point of view
between the switching activity VCD collected in the simulation and
the power trace acquired by the oscilloscope during the prototype execution.
On the other hand, it makes it possible to collect the VCD solely for
the time window of interest to the SCA analysis,
thus limiting the simulation's execution time.
The latter is a particularly critical aspect due to the need to
perform a timing simulation of a post-route netlist to have
maximum correspondence with the design prototyped on the board and
thus collect meaningful switching activity statistics.

\subsection{Experimental analysis}
\label{ssec:exp_analysis}
The experimental analysis applies to the computing platform, configured as
described in Section~\ref{ssec:exp_setup}, the SCA countermeasure and attack
techniques from the state of the art included in the framework,
quantitatively assessing their effectiveness.

\subsubsection{SCA countermeasure techniques}
\label{sssec:exp_defense}
The SCA countermeasures include microarchitecture-level ones,
such as clock frequency randomization, and
architecture- and OS-level ones, such as
the morphing and chaff techniques.

\paragraph*{Clock frequency randomization}
The DFS actuator can be used to continuously vary, in a random fashion,
the frequency of the clock signal fed to the CPU, as previously described
in Section~\ref{ssec:platform_dfs}, with the purpose of
producing a distortion in the power consumption trace and
therefore reducing the information leakage~\cite{Galli_2024ICECS}.
The DFS, when in random reconfiguration mode, uses indeed
as its target frequencies the values generated by the internal TRNG,
rather than setting them through the debug infrastructure.
The DFS actuator produces a clock signal that can
be reconfigured to up to 1024 different frequencies
every few tens of microseconds and
without any glitching and gating effects,
enabling a large variability and ensuring
the effectiveness of the countermeasure.

\begin{algorithm}[!t]
	\caption{Chaff SCA countermeasure~\cite{Agosta_2015DAC}.}
	\label{alg:chaff}
	\begin{algorithmic}[1]
		\Function{ChaffEncrypt\ }{cipher, key, ptx, numChaff}
			\State mainThread $\gets$ C\scriptsize{REATE} \normalsize (cipher, key, ptx)
			\State chaffKey $\gets$ G\scriptsize{ENERATE}\normalsize{C}\scriptsize{HAFF}\normalsize{K}\scriptsize{EYS} \normalsize (key, numChaff)
			\For {i in 1 to numChaff}
				\State chaffThreads[i] $\gets$ C\scriptsize{REATE} \normalsize (cipher, chaffKey[i], ptx)
			\EndFor
			\State S\scriptsize{TART}\normalsize{R}\scriptsize{ANDOM}\normalsize{S}\scriptsize{CHEDULER} \normalsize (mainThread, chaffThreads)
			\While{I\scriptsize{S}\normalsize{R}\scriptsize{UNNING} \normalsize (mainThread)}
				\State W\scriptsize{AIT} \normalsize ()
			\EndWhile
			\For {i in 1 to numChaff}
				\State K\scriptsize{ILL} \normalsize (chaffThreads[i])
			\EndFor
		\EndFunction
	\end{algorithmic}
\end{algorithm}

\paragraph*{Morphing}
The morphing countermeasure leverages the TRNG to randomly
modify the execution of a cryptographic operation~\cite{Agosta_2015TCAD}.
For example, in the AES use case considered in the experimental evaluation,
morphing targets the AddRoundKey and SubBytes steps of the AES cryptosystem.
Each AddRoundKey execution employs a randomly selected version of
the XOR operation, chosen among 8 equivalent implementations with
different power consumption profiles.
SubBytes executions are instead morphed by using S-Boxes
that are masked by random values and periodically refreshed.

\paragraph*{Chaff}
The TRNG component and the support for software multithreading
provided by FreeRTOS enable the adoption of
the chaff approach, described in~\cite{Agosta_2015DAC}
and whose pseudocode is listed in Algorithm~\ref{alg:chaff}.
The latter technique executes, concurrently to a thread for
the actual instance of a cipher~(line 2 in Algorithm~\ref{alg:chaff}),
e.g., the encryption of a plaintext with a certain key,
a set of additional threads performing the same cryptographic operation
on the same input but with different keys,
properly generated to be correlated with the original one~(lines 3--5).
All the threads are started and executed in parallel, according to
a random scheduler that leverages the SoC's TRNG~(line 6),
until the thread performing the actual cryptographic operation has completed~(lines 7--10).

\subsubsection{SCA attack techniques}
\label{sssec:exp_attack}
The SCA attacks provided out of the box by the framework include
\begin{enumerate*}[label=\textit{\arabic*)}]
	\item a \emph{CPA} attack~\cite{Brier_2004CHES},
	that extracts the secret key of the target cryptosystem executions by
	correlating their power signature with their operating behavior,
	performed in the experimental campaign on up to 1 million power traces,
	\item a \emph{template} profiled attack~\cite{Chari_2002CHES}, that leverages the Bayes' theorem
	to estimate the probability of a key given its multivariate normal distribution,
	carried out by using 1024 traces for each of the 256 possible key bytes
	during both the profiling and attack phases, and
	\item an ML-based attack leveraging convolutional neural networks~(\emph{CNN}s)~\cite{Benadjila_2020JCE},
	that learn a recurrent pattern of the leakage to predict the secret key and
	whose training and inference are carried out on the same dataset as the template attack. 
\end{enumerate*}

\subsubsection{SCA security assessment}
\label{sssec:exp_results}
The experimental analysis assesses first the vulnerability of the computing platform
to the previously described attacks while executing the AES-128 cryptosystem.
We make use of the constant-time, S-box-based AES-128 C implementation from
the widely adopted OpenSSL~\cite{openSSL} cryptography suite.
In particular, we evaluate whether each attack,
specifically targeting the SubBytes step of the first AES round,
succeeds or fails when executing
the plain AES cryptosystem without any protection,
a software-masked version of AES,
and plain AES protected by the clock frequency randomization,
morphing, and chaff SCA hiding countermeasures.

\begin{table}[t]
	\centering
	\caption{Quality metrics obtained by attacks to AES execution on the experimental platform
		when implementing different countermeasures.
		Undefined values are denoted by --.}
	\begin{tabular}{llrrr}
		\toprule
		                                  &                         &                                    \multicolumn{3}{c}{\textbf{Attack}}                                    \\ \cmidrule(lr){3-5}
		\textbf{Countermeasure}           & \textbf{Quality metric} &                      \textbf{CPA} &                 \textbf{Template} &                      \textbf{CNN} \\ \midrule
		\multirow{4}{*}{None}             & Guessing entropy        &                                 1 &                                 1 &                                 1 \\
		                                  & Guessing distance       &                                -- &                              0.86 &                              0.77 \\
		                                  & Success rate            & \textcolor{Green}{\textbf{100\%}} & \textcolor{Green}{\textbf{100\%}} & \textcolor{Green}{\textbf{100\%}} \\
		                                  & Number of traces        &                               180 &                                 3 &                                10 \\ \cmidrule(lr){1-5}
		\multirow{4}{*}{Software masking} & Guessing entropy        &                            108.75 &                            153.22 &                              1.06 \\
		                                  & Guessing distance       &                                -- &                             -0.57 &                              0.45 \\
		                                  & Success rate            &     \textcolor{Red}{\textbf{0\%}} &     \textcolor{Red}{\textbf{0\%}} &  \textcolor{Green}{\textbf{98\%}} \\
		                                  & Number of traces        &                                -- &                                -- &                                -- \\ \cmidrule(lr){1-5}
		\multirow{3}{*}{Clock frequency}  & Guessing entropy        &                            112.75 &                               127 &                             72.20 \\
		\multirow{3}{*}{randomization}    & Guessing distance       &                                -- &                             -0.52 &                             -0.31 \\
		                                  & Success rate            &     \textcolor{Red}{\textbf{0\%}} &     \textcolor{Red}{\textbf{0\%}} &     \textcolor{Red}{\textbf{6\%}} \\
		                                  & Number of traces        &                                -- &                                -- &                                -- \\ \cmidrule(lr){1-5}
		\multirow{4}{*}{Morphing}         & Guessing entropy        &                                 1 &                                 1 &                                 1 \\
		                                  & Guessing distance       &                                -- &                              0.38 &                              0.81 \\
		                                  & Success rate            & \textcolor{Green}{\textbf{100\%}} & \textcolor{Green}{\textbf{100\%}} & \textcolor{Green}{\textbf{100\%}} \\
		                                  & Number of traces        &                              3072 &                               580 &                                 8 \\ \cmidrule(lr){1-5}
		\multirow{4}{*}{Chaff}            & Guessing entropy        &                            105.25 &                               135 &                               122 \\
		                                  & Guessing distance       &                                -- &                             -0.52 &                             -0.48 \\
		                                  & Success rate            &     \textcolor{Red}{\textbf{0\%}} &     \textcolor{Red}{\textbf{0\%}} &     \textcolor{Red}{\textbf{0\%}} \\
		                                  & Number of traces        &                                -- &                                -- &                                -- \\ \bottomrule
	\end{tabular}
	\label{tab:aes_attack_results_numbers}
\end{table}

The attacks' effectiveness is evaluated according to
\begin{enumerate*}[label=\textit{\arabic*)}]
	\item the \emph{guessing entropy}, defined as
	the average rank position of the correct key among all possible key guesses,
	\item the \emph{guessing distance}, which represents the normalized probability distance
	between the correct key and the first-ranked non-correct one,
	\item the \emph{success rate}, i.e., the percentage of attacks that succeed in delivering the secret key, and
	\item the minimum \emph{number of traces} required to obtain a prediction that is always correct.
\end{enumerate*}
Table~\ref{tab:aes_attack_results_numbers} lists the values of
such four quality metrics obtained for the various countermeasure-attack
combinations considered in the experimental campaign.
In particular, the attacks that always succeed,
i.e., with guessing entropy equal to 1 and 100\% success rate, respectively,
list the corresponding number of traces employed to achieve such a perfect attack outcome.
The execution of the plain AES application is successfully attacked
by all three considered techniques, while the masked version is broken
only by the CNN attack.
The clock frequency randomization and chaff countermeasures prove instead to be
effective against the three SCA attacks, whereas applying morphing
does not protect the computing platform from any of them.

More in detail, clock frequency randomization and chaff are the most effective countermeasures
against the proposed SCA attacks, as they achieve a high guessing entropy and
a success rate close to 0\% in every scenario.
The morphing countermeasure is instead shown to be more vulnerable, as every technique can break it.
The CNN attack can isolate the correct key with a guessing distance of 0.81,
using only eight traces, while CPA and template attacks are less effective though successful.
Guessing distance values are notably undefined for the CPA attack, since the latter is not a 
probabilistic attack. Guessing distance measures indeed a probability distance and,
thus, it is suitable only for attacks such as template and CNN ones
that compute the probability of a key guess being correct.

\begin{figure}
	\begin{minipage}[b]{0.98\columnwidth}
		\centering
		\begin{subfigure}[b]{\columnwidth}
			\centering
			\includegraphics[width=\textwidth]{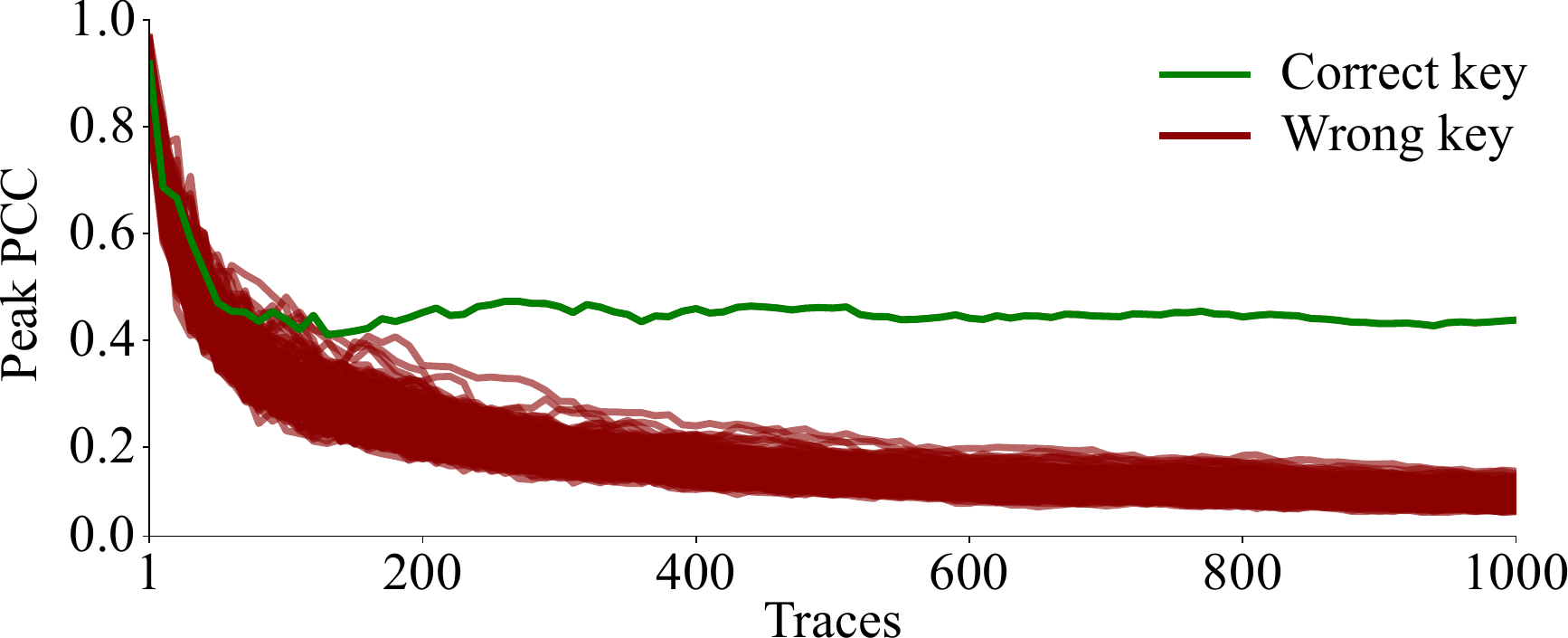}
			\caption{Pearson correlation coefficient (PCC) of CPA attack}
			\label{sfig:cpa_results}
			\vspace{0.2cm}
		\end{subfigure}
		\begin{subfigure}[b]{\columnwidth}
			\centering
			\includegraphics[width=\textwidth]{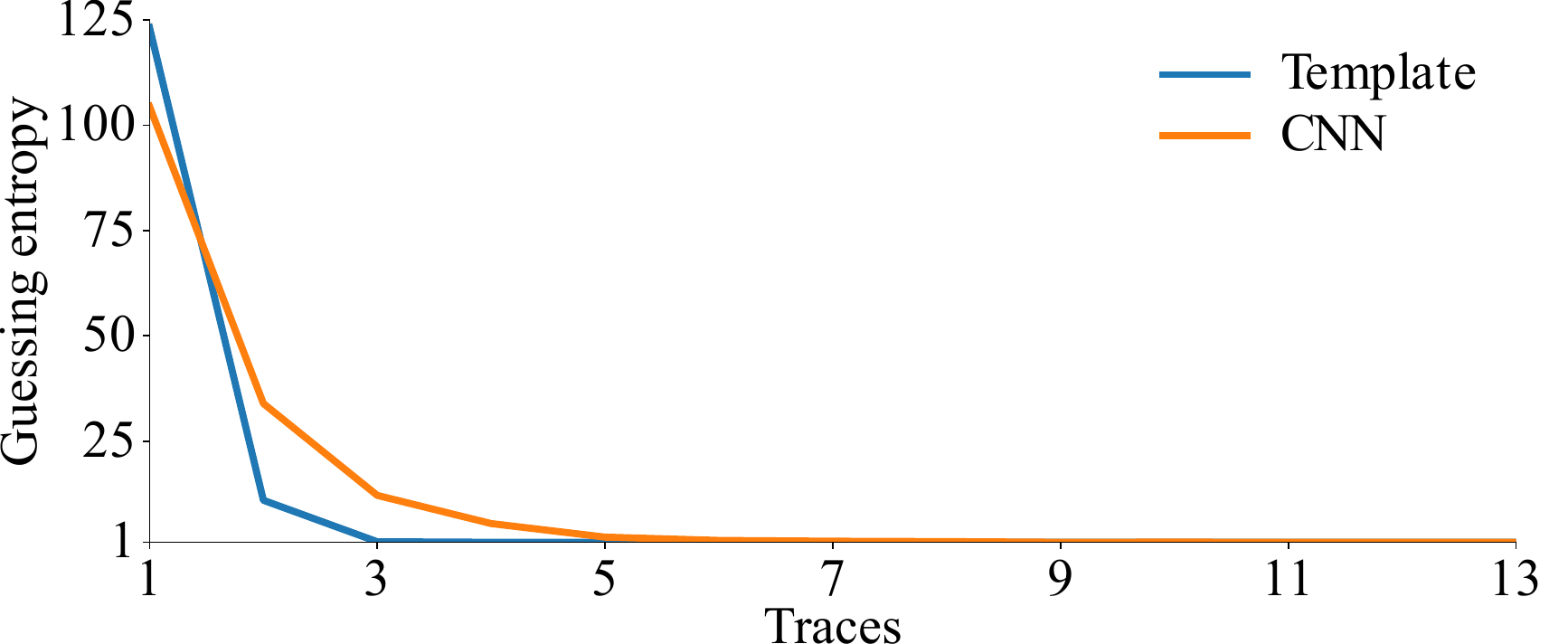}
			\caption{Guessing entropy of template and CNN attacks}
			\label{sfig:ge_results}
		\end{subfigure}
	\end{minipage}
	\caption{SCA attacks against unprotected AES.}
	\label{fig:attack_results}
\end{figure}

Figure~\ref{fig:attack_results} depicts how the SCA attacks against the execution of
plain AES evolve as the number of traces increases when applying no SCA countermeasure.
The CPA attack, as shown in Figure~\ref{sfig:cpa_results},
requires 180 traces to correctly identify the correct key.
More traces are needed to reduce the correlation coefficient of the wrong key guesses, drawn in red,
whereas the Pearson correlation coefficient~(PCC) of the correct key, drawn in green,
remains steadily around 0.43.
Figure~\ref{sfig:ge_results} demonstrates instead how the guessing entropy
improves in the template and CNN attacks as more traces get analyzed and
how it stabilizes to 1 after 3 and 10 traces have been processed, respectively.

\begin{table}[t]
	\centering
	\caption{Quality metrics obtained by template attacks to Camellia, Clefia, and Seed execution
	    when implementing different countermeasures.
		Undefined values are denoted by --.}
	\begin{tabular}{llrrr}
		\toprule
		                                 &                         &                                 \multicolumn{3}{c}{\textbf{Cryptosystem}}                                 \\ \cmidrule(lr){3-5}
		\textbf{Countermeasure}          & \textbf{Quality metric} &                 \textbf{Camellia} &                   \textbf{Clefia} &                     \textbf{Seed} \\ \midrule
		\multirow{4}{*}{None}            & Guessing entropy        &                                 1 &                                 1 &                                 1 \\
		                                 & Guessing distance       &                              0.28 &                              0.61 &                              0.23 \\
		                                 & Success rate            & \textcolor{Green}{\textbf{100\%}} & \textcolor{Green}{\textbf{100\%}} & \textcolor{Green}{\textbf{100\%}} \\
		                                 & Number of traces        &                              1202 &                                52 &                              1021 \\ \cmidrule(lr){1-5}
		\multirow{3}{*}{Clock frequency} & Guessing entropy        &                            133.09 &                            105.53 &                            132.63 \\
		\multirow{3}{*}{randomization}   & Guessing distance       &                             -0.49 &                             -0.45 &                             -0.50 \\
		                                 & Success rate            &     \textcolor{Red}{\textbf{0\%}} &     \textcolor{Red}{\textbf{0\%}} &     \textcolor{Red}{\textbf{0\%}} \\
		                                 & Number of traces        &                                -- &                                -- &                                -- \\ \cmidrule(lr){1-5}
		\multirow{4}{*}{Morphing}        & Guessing entropy        &                                 9 &                                 1 &                             55.75 \\
		                                 & Guessing distance       &                             -0.06 &                              0.31 &                             -0.31 \\
		                                 & Success rate            & \textcolor{Orange}{\textbf{40\%}} & \textcolor{Green}{\textbf{100\%}} &     \textcolor{Red}{\textbf{3\%}} \\
		                                 & Number of traces        &                                -- &                              1379 &                                -- \\ \cmidrule(lr){1-5}
		\multirow{4}{*}{Chaff}           & Guessing entropy        &                            148.06 &                            119.72 &                            133.84 \\
		                                 & Guessing distance       &                             -0.56 &                             -0.48 &                             -0.56 \\
		                                 & Success rate            &     \textcolor{Red}{\textbf{0\%}} &     \textcolor{Red}{\textbf{0\%}} &     \textcolor{Red}{\textbf{0\%}} \\
		                                 & Number of traces        &                                -- &                                -- &                                -- \\ \bottomrule
	\end{tabular}
	\label{tab:other_attack_results_numbers}
\end{table}

To further demonstrate the flexibility of the proposed hardware-software
framework, Table~\ref{tab:other_attack_results_numbers} reports the results
of the template attack when the computing platform executes three additional
cryptographic applications from the OpenSSL~\cite{openSSL} suite, namely, the
Camellia, Clefia, and Seed block ciphers. For each evaluated cryptosystem,
results are reported considering no countermeasures, the use of clock frequency
randomization, the use of morphing and the use of chaffing. 
The reported quality metrics to assess the side-channel vulnerability refer to
the SubBytes operation in the first round of each cryptosystem.

The results show that applying morphing to protect Clefia is ineffective against
template attacks, whereas it is moderately and highly effective in the Camellia
and Seed use cases, respectively. Conversely, the clock frequency randomization
and chaff countermeasures are shown to successfully thwart the template attack
when applied during the execution of all three cryptographic applications.

\section{Conclusions}
\label{sec:conclusions}
This manuscript introduced a novel open-source framework for research on SCA targeting
FPGA-based IoT-class computing platforms.
The framework includes a RISC-V-based IoT-class SoC that features
an ad-hoc debug infrastructure to maximize the observability and controllability of
the computing platform and thus simplify the execution of SCA attacks,
as well as a DFS actuator, a TRNG, and a timer that provide support for
a set of state-of-the-art SCA countermeasures available out of the box.
A complete automated flow encompasses the configuration of the SoC,
the execution of target applications and corresponding collection of side-channel information,
and the analysis to identify eventual SCA vulnerabilities and
pinpoint the sources of side-channel information leakage.

The user is encouraged and empowered to
expand the capabilities of the hardware-software infrastructure
and support novel SCA attacks and countermeasures by
the open-source nature of the framework,
its adoption of standard languages for both its hardware and software components,
and the usage of widely available devices and tools.

Future developments foresee the addition of a dual-core
CPU architecture and of accelerators specifically dedicated to cryptography purposes.
Such optional features, limited in terms of additional resources to
still deliver a lightweight IoT-class computing platform, will not compromise
the observability and controllability of the system for SCA,
which are essential for detecting side-channel leakage and its sources.
In addition, we plan to extend our framework to support research on
fault attacks and countermeasures against the latter.

\bibliographystyle{IEEEtran}
\bibliography{2025_TC}

\def\PhotoSkip{\vspace*{-1.01cm}}

\PhotoSkip
\begin{IEEEbiography}[{\includegraphics[width=1in,height=1.25in,clip,keepaspectratio]{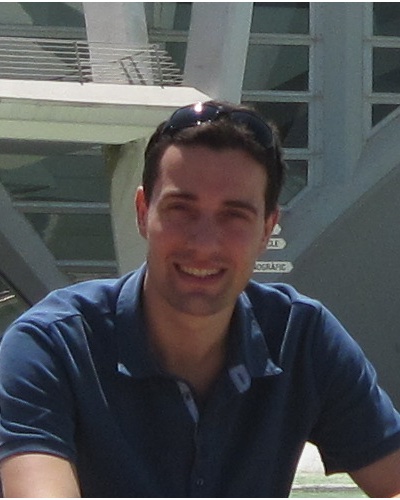}}]{Davide Zoni,}
	PhD, is associate professor at Politecnico di Milano, Italy.
	He published more than 50 papers in journals and conference proceedings.
	His research interests include RTL design and verification of single- and multi-cores at the edge
	with emphasis on low power, hardware security, and deep learning.
	He filed two patents on cyber-security, and he is co-founder at Blue Signals, a
	spin-off of Politecnico di Milano.
\end{IEEEbiography}
\PhotoSkip
\begin{IEEEbiography}[{\includegraphics[width=1in,height=1.25in,clip,keepaspectratio]{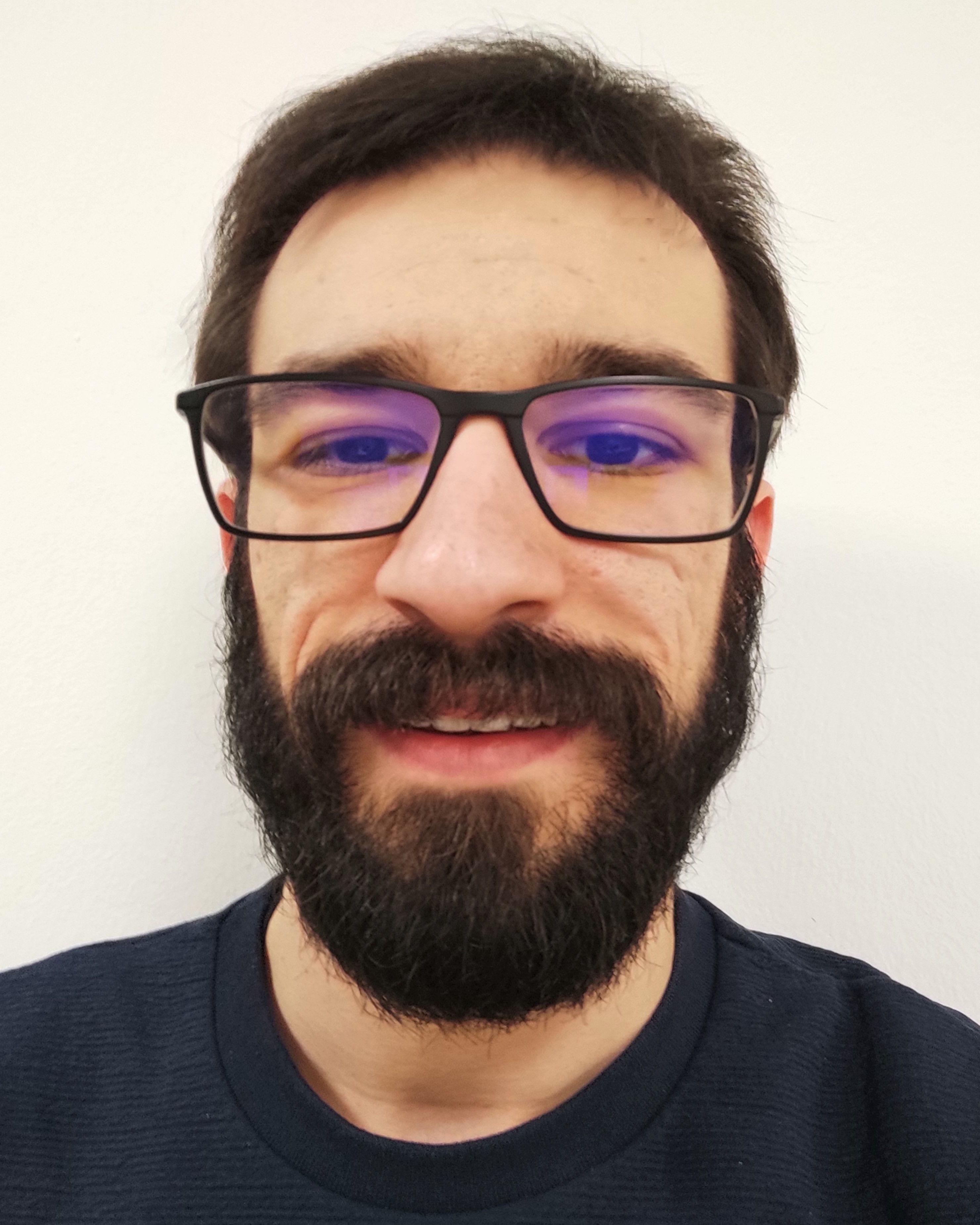}}]{Andrea Galimberti,}
	PhD, is a research fellow at Politecnico di Milano.
	He received his M.Sc. degree in Computer Science and Engineering in 2019 and
	his PhD degree in Information Technology in 2023, both from Politecnico di Milano.
	His research has explored hardware acceleration of post-quantum cryptography and
	architectures for mixed-precision computing integrating floating-point and fixed-point arithmetic.
	His current interests focus on designing multi-core architectures and accelerators for deep learning.
\end{IEEEbiography}
\PhotoSkip
\begin{IEEEbiography}[{\includegraphics[width=1in,height=1.25in,clip,keepaspectratio]{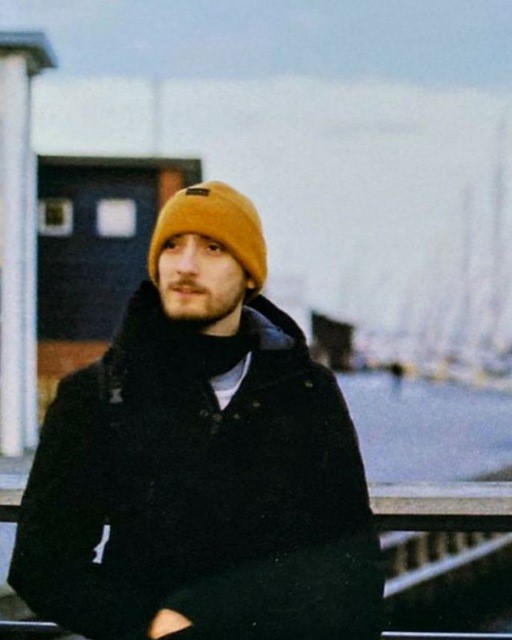}}]{Davide Galli,}
	MSc, is a PhD student at Politecnico di Milano, Italy.
	He received the B.Sc. in Ingegneria Informatica and the M.Sc.
	in Computer Science and Engineering at Politecnico di Milano in 2019 and
	2022, respectively. Starting from his M.Sc. thesis on true random number
	generators on FPGA, his research interests are mainly on hardware security
	and side-channel analysis.
\end{IEEEbiography}

\vfill

\end{document}